\documentclass[12pt, preprint]{aastex}

\usepackage{amsmath}
\usepackage{amssymb}

\newcommand{\dd}{\mathrm{d}}
\newcommand{\pc}{\mathrm{pc}}

\shorttitle{Black Hole Growth}
\shortauthors{MacMillan \& Henriksen}

\begin{document}

\title{Black Hole Growth in Dark Matter and the $M_{bh} - \sigma$ Relation}

\author{J. D. MacMillan and R. N. Henriksen}
\affil{Department of Physics, Queen's University, Kingston, Ontario, Canada, K7L 3N6}
\email{macmilla@astro.queensu.ca, henriksn@astro.queensu.ca}

\begin{abstract}
In this article we consider the growth of seed black holes immersed in dark matter halos. We first investigate the adiabatic growth in various initial distribution functions (isothermal, power law, and NFW) and find the resulting density, radial velocity, and anisotropy profiles. In addition we estimate the growth rate for a given black hole mass in the corresponding adiabatically modified dark matter distribution function. Only in the isothermal case is there a convincing black hole mass-age relation. By calculating the line of sight velocity dispersion for the various cases as a function of the black hole mass, we find the predicted adiabatic $M_{bh}-\sigma$ relation; this never approaches the recently observed power law. We conclude by abandoning adiabaticity, suggesting that the black hole grows proportionally to the dark matter halo itself on a dynamic time scale. This allows us to relate the observed $M_{bh}-\sigma$ relation to the cosmological power spectrum on galactic scales by using dimensional scaling arguments.     
\end{abstract}

\keywords{black hole physics---galaxies: evolution---galaxies: halos---galaxies: nuclei---dark matter}


\section{INTRODUCTION}

Recent results by \citet{geb00} and by \citet{fer00} that establish a strong correlation between central black hole mass $M_{bh}$ and the velocity dispersion $\sigma_e$, measured at $r_e/8$, indicate that the central black hole is intimately related to the dynamical structure of the galaxy. An earlier result by \citet{mag98} relates $M_{bh}$ linearly to the mass of the bulge, which suggests a similar conclusion, while Merrifield, Forbes, \& Terlevich (2000) establish a link between $M_{bh}$ and the age of the stellar system: a massive central black hole seems to grow over a period of a Gyr or so. 

These results have already stimulated the emission of various theories that create a feed-back mechanism between bulge star formation and accretion of gas onto the black hole \citep{bur01}. These theories generally tend to establish the desired correlations but at the cost of some rather complicated physics simply, perhaps oversimply, described. 

The object of the present paper is to first revisit the correlations established by the adiabatic growth of a black hole in a galaxy, since this process is relatively free of physical assumptions once the initial distribution function (DF) is chosen. This part of the work is very much in the spirit of \citet{mar99}, who examined the effect of the black hole on the fundamental plane relations, but did not consider explicitly the $M_{bh} - \sigma$ relation. We choose distribution functions moreover that are appropriate for collisionless dark matter halos, the philosophy being that these are the dynamically dominant components of massive galaxies and should therefore dictate the observed velocities. These are chosen to be an isothermal or Gaussian DF (to test our code against previous work and because this may be the maximum entropy state according to \citet{nak00}), an isotropic steady-state power law DF found by \citet{hen95} for comparison purposes, and the DF that corresponds to the NFW density profile, as approximated by \citet{wid00}, as the best measured approximation to a dark-matter halo. 

Such an approach does not, of course, explain the origin of the black hole, but rather yields only the perturbed DF that is created by its adiabatically established presence. An earlier attempt to grow the black hole during the formation of the galaxy \citep{ost00} relied on dissipative dark matter which is fraught unfortunately with badly known parameters. In the present approach the black hole growth is limited to the particle flux across the event horizon of a seed black hole that is peculiar to the initial or adiabatically modified DF. 

The adiabatic approach fails to yield either the correct form or extent of the observed correlations between the black hole mass and the modified galaxy. We therefore suggest a possible explanation of the correlations based on a non-adiabatic process of black hole growth on the formation timescale of the dark matter halo. The argument is essentially dimensional at this stage and must be examined numerically in greater detail.

In section 2 we review the adiabatic growth approximation and verify our code with the isothermal DF. In section 3 we give the results in the power law and NFW distribution functions. In section 4 we discuss the resulting $M_{bh}$ versus $\sigma_e$ in the various distribution functions and show the extent of the black hole influence in the galaxy. In section 5 we give our dimensional derivation of the observed relations based on a self-similar (but not necessarily spherically symmetric) growth of black hole and dark halo. Finally we give our conclusions.       
     

\section{ADIABATIC GROWTH}

\subsection{History}

We consider the slow growth of an initially small seed black hole (BH) located at the centre of a collisionless, spherical system of stars and dark matter.  Using an algorithm that makes use of the fact that, under these circumstances, the radial and transverse actions are conserved, the final state of the system containing a supermassive BH can be calculated.

This technique was first suggested by \citet{pee72}, who used it to show that an adiabatic cusp with $\varrho \sim R^{-3/2}$ would form in an isothermal sphere.  This work was confirmed numerically by \citet{you80}.  His algorithm is used here essentially unchanged.  \citet{lee87} used Young's algorithm to explore the more general case of a stellar system with a net rotation, although they treat the potential as spherically symmetric rather than axisymmetric.  They found that, while the rotation to dispersion ratio $v/\sigma$ was larger after adiabatic growth, the gain was not enough to match the observed values.  A variety of systems were examined in detail by \citet{qui95}, and \citet{mar99} used their code to explain a number of observational properties of galaxies.  Finally, the adiabatic growth model was used by \citet{gon99} to examine how dark matter annhilations in the cusp that would form in our own galaxy could be used as a probe of the dynamics or of the nature of dark matter.  They suggest that searching for a neutrino signal from the central density spike (the neutrino flux increases with an increase in cusp slope) could set upper bounds to the dark matter cusp in our galaxy, or, alternatively, clarify the nature of dark matter.

Although these studies serve to demonstrate the utility and topicality of the technique, they have not studied explicitly the relation between dark matter distribution functions and the central black hole mass that is our concern below.  

\subsection{Units}

For convenience, we use here dimensionless units, characterized by the variables
\begin{equation}
\begin{array}{lll}
R & = & r/r_0, \\
\varrho & = & \rho/\rho_0, \\
\varepsilon & = & (E - \Phi_0)/(4 \pi G \rho_0 r_0^2), \\
\psi & = & (\Phi - \Phi_0)/ (4 \pi G \rho_0 r_0^2), \\
j & = & J / [4 \pi G \rho_0 r_0^4]^{1/2}, \\
F & = & (4 \pi G)^{3/2}  \rho^{1/2} r_0^3 f, \\
m &  = & M / (4 \pi \rho_0 r_0^3), \\
\end{array}
\end{equation}
The fiducial quantities $r_0$ and $\rho_0$ are usually taken to be the core radius of the galaxy, similar to the usual definition of the King radius, and the density at this radius (where the surface density is about one half of its central value in a fitted isothermal sphere).

\subsection{Theory}

The initial system, containing the negligible mass BH, can be described by a distribution function $F(\varepsilon, j)$, and the density of the system can be calculated from
\begin{equation}
\label{eq:density}
\varrho(R) = 4 \pi \int_{\psi(R)}^{\psi(\infty)} \dd \varepsilon \int_0^{j_{max}} \frac{j \dd j}{R^2 V_R} F(\varepsilon, j),
\end{equation}
where the radial velocity is expressed as $V_R = [2(\varepsilon-\psi) - j^2/R^2]^{1/2}$ and $j_{max} = [2R^2 (\varepsilon-\psi)]^{1/2}$ is the maximum angular momentum.  The potential of the system can then be calculated from Poisson's equation,
\begin{equation}
\frac{1}{R^2} \frac{\dd}{\dd R} \left( R^2 \frac{\dd \psi}{\dd R} \right) = \varrho.
\end{equation}
The integral form of Poisson's equation is often more convenient, however.  We take the reference potential to be zero at the centre of the system and we suppose the halo mass to tend to zero there, which is true for a density profile less steep than $R^{-3}$. Hence $\psi = \dd \psi / \dd R = 0$ for $R = 0$, and we have
\begin{equation}
\label{eq:potential}
\psi(R) = \int_0^R \varrho(s) s \dd s - \frac{m(R)}{R},
\end{equation}
where the total mass of the system is
\begin{equation}
m(R) = \int_0^r \varrho(s) s^2 \dd s + m_{bh},
\end{equation}
and $m_{bh}$ is the mass of a central object, if one exists.

Now, if the growth of the central seed black hole is much slower than the dynamical time of the system, the radial and transverse actions of the system are conserved.  These so-called adiabatic invariants are given by \citep{you80}
\begin{equation}
\label{eq:radial}
\begin{array}{lll}
i_R & = & 2 \int_{R_-}^{R_+} \, V_R \dd R \\
& = & 2 \int_{R_-}^{R_+} \, \sqrt{2(\varepsilon-\psi) - j^2/R^2} \, \dd R,
\end{array}
\end{equation}
where $R_\pm$ are the turning points of the orbit (zeros of the integrand), and
\begin{equation}
\begin{array}{lll}
 i_T & = & \int_0^{2\pi} V_T \dd (R\theta) \\
& = & 2\pi j,
\end{array}
\end{equation}
so that the angular momentum is an action variable.

The adiabatic growth framework is then as follows.  A particle of the initial system with energy $\varepsilon$ and angular momentum $j$ will have a different energy $\varepsilon^{*}$ (but identical angular momentum $j$) after the BH has grown.  However, its radial action $i_R$ will be the same before and after growth;  thus, it is possible to invert $i_R^{*}(\varepsilon^{*}, j)$ to get $\varepsilon$ as a function of $(\varepsilon^{*}, j)$.  The final DF is then found from
$F^{*}(\varepsilon^{*}, j) = F(\varepsilon(\varepsilon^{*}, j), j)$.  

This is implemented in an algorithm similar to that used first by \citet{you80}.  See Appendix A for details.

\subsection{The Isothermal Sphere}

We apply the adiabatic growth framework first to the isothermal sphere, characterized by the Gaussian DF
\begin{equation}
F(\varepsilon) = (2 \pi)^{-3/2} \, e^{-\varepsilon}.
\end{equation}

This is a well-studied distribution, and was Young's (1980) initial system.  We present the adiabatic results for the isothermal sphere here mainly as a check that the code developed gave the same results that \cite{you80} found. There are in fact reasons to believe \citep{nak00, hen01} that this is ultimately the DF of interest, especially near the centre of the system. Of course the system may not be fully relaxed when the black hole is growing. 

\begin{figure}
\epsscale{0.8}
\plotone{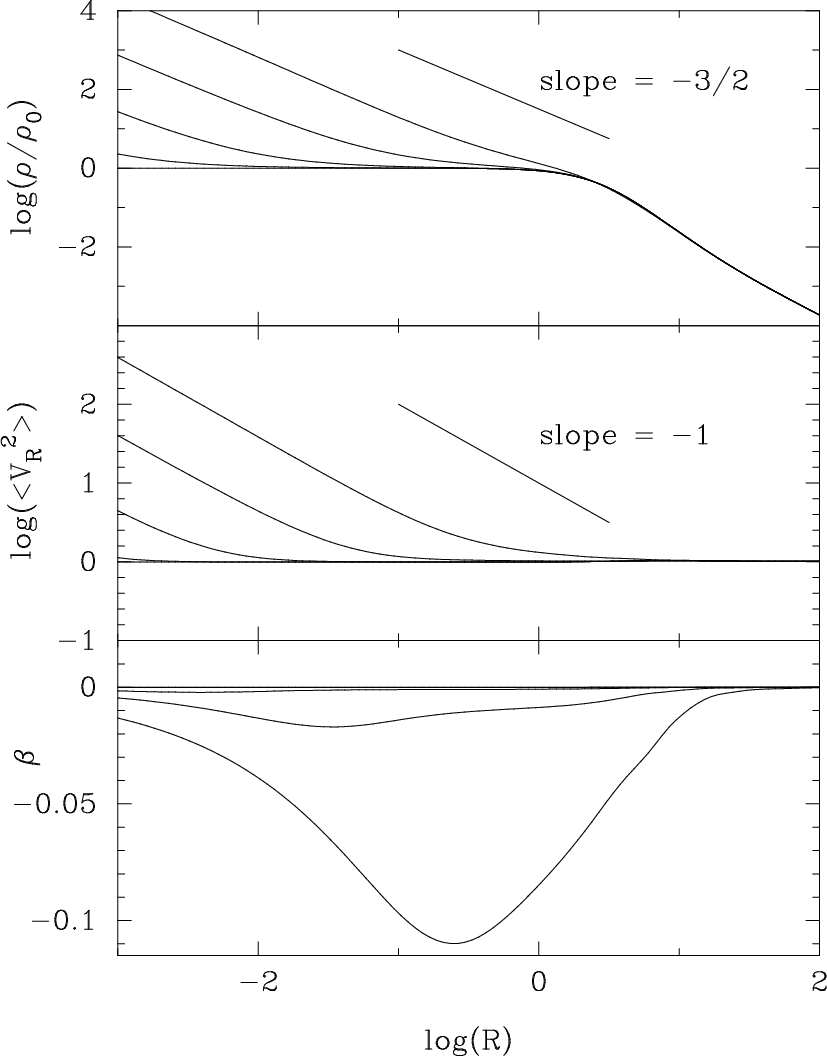}
\caption{Adiabatic growth in the isothermal sphere for masses $m_{bh} = 0$, 0.001, 0.01, 0.1, and 1.0, with the mass increasing from bottom to top in the top two panels, and top to bottom in the last panel.  Density is shown in the top frame, the radial velocity in the middle, and the anisotropy paramter $\beta$ in the bottom frame.}
\label{fig:iso}
\end{figure}

Results are shown in Figure \ref{fig:iso} for black hole masses from zero through $0.001$, $0.01$, $0.1$ and $1.0$ as labelled by  increasing effect.  These dimensionless masses correspond to physical masses in the range of about $10^7$ to $10^{10}$ $M_{\odot}$ if we use the values of $\rho_0$ and $r_0$ given in section \ref{sec:growth}.

We find the same characteristic density cusp of $R^{-3/2}$ as obtained by \citet{you80}, and the radial velocity profiles are also similar. Although \citet{you80} did not show results for the anisotropy parameter, defined as $\beta = 1 - <V_T^2> / <2 V_R^2>$, our results do agree with the prediction of \citet{qui95} for distributions of this type. We observe that there is at most about a $10$\% anisotropy in favour of tangential  motion, due to the increasing binding energy of a particle and the  resultant decrease in eccentricity at constant angular momentum. Moreover the perturbation extends as far as the core radius  only when the mass is comparable to that of the core, as is to be expected. These results are the main test of our program. 


\section{CUSPY DARK MATTER HALOS }

We explore here the adiabatic growth of a central black hole in a collisionless dark matter halo that possesses a central cusp.  For initial systems, we choose two very different starting points.  The first is an isotropic self-similar system, meant to represent the final state of a halo undergoing self-similar relaxation \citep{hen95}.  The other system is the NFW system of \citet{nav96}, which fits a wide range of dark matter halo sizes.

\subsection{Self-similar Distribution}

The self-similar system is described by the DF 
\begin{equation}
F(\varepsilon) = F_0 |\varepsilon|^{-\frac{3 \delta - 1}{2 \delta - 2}},
\end{equation}
where $\delta$ is a free parameter ($2/3 < \delta < 1,\ \delta > 1$) that essentially controls the logarithmic slope of the initial density and potential, given by
\begin{equation}
\varrho = R^{-2/\delta} \quad \mathrm{and}
\end{equation}
\begin{equation}
\psi = R^{2 - 2/\delta}.
\end{equation}
The constant $F_0$ depends only on the value of $\delta$, and can be solved for in terms of the core radius $r_0$ and the density $\rho_0$ by using equations (\ref{eq:density}) and (\ref{eq:potential}).

\begin{figure}
\epsscale{0.8}
\plotone{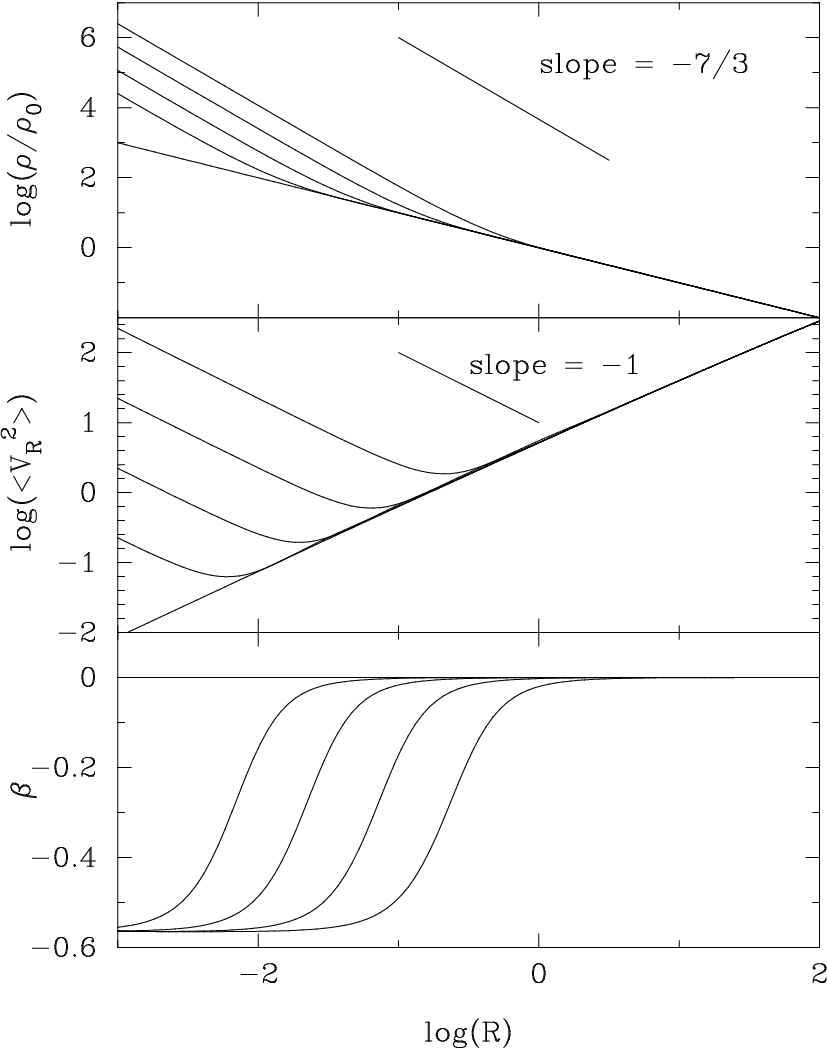}
\caption{Adiabatic growth in the self-similar system with parameter $\delta = 2$ and masses $m_{bh} = 0$, 0.001, 0.01, 0.1, and 1.0, with the mass increasing from bottom to top in the top two panels, and top to bottom in the last panel.  Density is shown in the top frame, the radial velocity in the middle, and the anisotropy paramter $\beta$ in the bottom frame.}
\label{fig:self_2}
\end{figure}

\begin{figure}
\epsscale{0.8}
\plotone{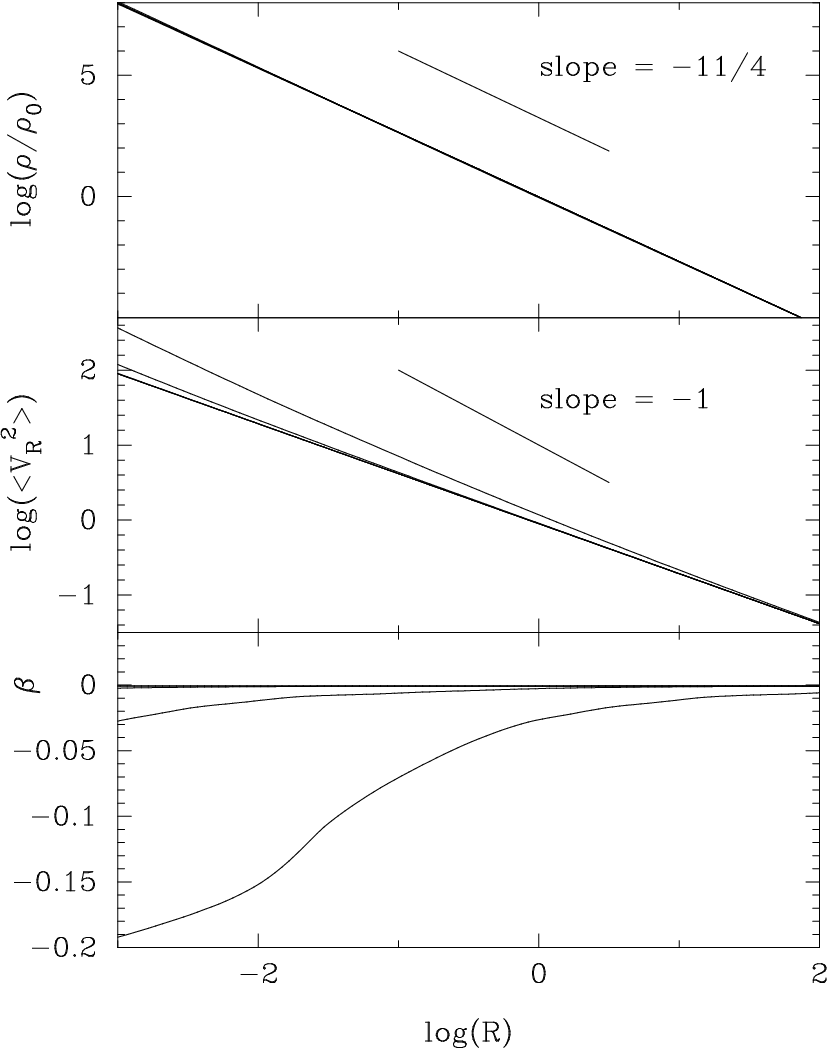}
\caption{Adiabatic growth in the self-similar system with parameter $\delta = 3/4$ and masses $m_{bh} = 0$, 0.001, 0.01, 0.1, and 1.0, with the mass increasing from bottom to top in the top two panels, and top to bottom in the last panel. Density is shown in the top frame, the radial velocity in the middle, and the anisotropy paramter $\beta$ in the bottom frame.}
\label{fig:self_0.75}
\end{figure}

Using this system as a starting point for the adiabatic algorithm discussed above leads to rather different results than are found for the isothermal sphere.  The results for two different choices of $\delta$ are given; Figure \ref{fig:self_2} is for the parameter $\delta = 2$, while  Figure \ref{fig:self_0.75} shows results with $\delta = 3/4$.

As can be in Figure \ref{fig:self_2}, the initial density profile of the self-similar system with $\delta = 2$ is a power law with a logarithmic slope of $-1$.  Adding an adiabatically grown BH to the system induces a steeper cusp region at small radii, and the outer radius at which this region begins depends of course on the final BH mass.  The new cusp slope is $-7/3$, greater than the density cusp induced in the isothermal sphere, but the difference between the before-and-after slopes are not as great here as in the isothermal case (a gain of $\sim 1.3$ here versus $1.5$ for the isothermal sphere). The major difference with the isothermal sphere however is best seen in the anisotropy parameter. The system becomes quite tangentially anisotropic at small radii.  

That the cusp slope will not increase as greatly for an initial system which already has a steep density cusp is shown best in Figure \ref{fig:self_0.75}, where the grown BH seems to have left no visible mark on the density slope (to within graphical resolution: the initial density slope is $-8/3$ while the measured final slope is $-11/4$).  This result is for $\delta = 3/4$.  Notice, however, that the BH growth has disturbed the radial velocity, which takes on the same $1/R$ shape (although only for the largest masses) seen previously in both the $\delta = 2$ self-similar system and the isothermal sphere.  That these systems all have similar velocities at small radii is simply indicative of their similar Keplerian potentials there.

\subsection{The NFW Profile}

The NFW system, given by the universal density profile
\begin{equation}
\varrho = \frac{1}{R(1+R)^2},
\end{equation}
is more complicated to describe than the other two systems discussed above, since it does not have an analytic distribution function.  We use here an analytic fit of a numerical calculation of the DF \citep{wid00}, which in our units is
\begin{equation}
F(\varepsilon) = F_1 (1-\varepsilon)^{3/2} \varepsilon^{-5/2} \left(- \frac{\mathrm{ln} (1 - \varepsilon)}{\varepsilon} \right)^q \, e^P,
\end{equation}
where $P = \sum_i p_i \varepsilon^i$ and the parameters are:
\begin{equation}
\begin{array}{lll}
F_1 & = & 9.1967 \times 10^{-2} \\
q & = & -2.7419 \\
p_1 & = & 0.3625 \\
p_2 & = & -0.5669 \\
p_3 & = & -0.0802 \\
p_4 & = & -0.4945. \\
\end{array}
\end{equation}

\begin{figure}
\epsscale{0.8}
\plotone{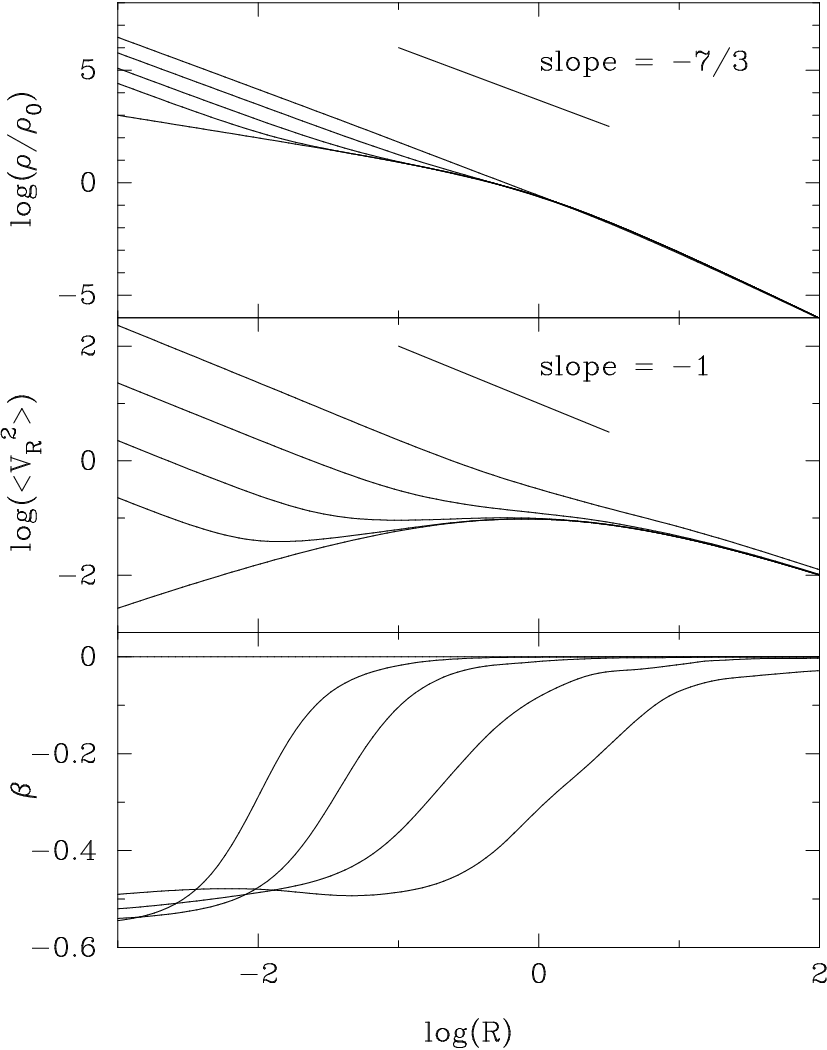}
\caption{Adiabatic growth in the NFW system with masses $m_{bh} = 0$, 0.001, 0.01, 0.1, and 1.0, with the mass increasing from bottom to top in the top two panels, and top to bottom in the last panel. Density is shown in the top frame, the radial velocity in the middle, and the anisotropy paramter $\beta$ in the bottom frame.}
\label{fig:nfw}
\end{figure}

The results for this system are given in Figure \ref{fig:nfw}.  As expected, the BH growth induces cusps in both the density and velocity.  The logarithmic slope for the density is $-7/3$, which is the same as for the self-similar system with $\delta = 2$; they also both began with the same slope ($\varrho \propto 1/R$) in the inner region where the BH disturbance is greatest.  The velocity cusp is the usual $1/R$. We notice that the velocity tangential anisotropy is also comparable to the system with $\delta=2$ at small radii for all black hole masses, and that the velocity perturbation can extend nearly to the core radius for $m_{bh}\ge 0.1$. 

\subsection{Growth Timescales}
\label{sec:growth}

It is of some interest to explore the rate at which a black hole would grow in the above dark matter distributions in view of the black hole mass galactic age relation reported by \citet{mer00}. Such growth represents an alternative to advection dominated accretion flow (ADAF; see \citet{nar00} and references therein) as a means to grow a black hole invisibly. We consider as an illustration the time to grow the black hole by a factor of ten in the various DFs. 

In order to calculate the timescale for the BH to grow by a factor of ten, we use the following expression for the growth of the BH as it accretes matter:
\begin{equation}
\frac{\dd M_{bh}}{\dd t} = 4 \pi r_x^2 \rho(r_x) <v_r>|_{r_x},
\end{equation}
where $r_x$ is the radius of the last stable orbit, $r_x = 3R_s$, and $R_s = 2 G M_{bh} / c^2$ is the Schwarschild radius of the BH.  In our units, and with $<v_r>$ given as an integral over the distribution function, this expression becomes
\begin{equation}
\label{eq:time}
\frac{\dd M_{bh}}{\dd t} = 8 \pi^2 (4 \pi G)^{1/2} \rho_0^{3/2} r_0^3  \int_{\psi_x}^{\psi(\infty)} \dd \varepsilon \int_0^{j_{max}} j \dd j F(\varepsilon, j),
\end{equation}
where $\psi_x = \psi(r_x)$ is the potential evaluated at the radius of the last stable orbit.  Note that this equation requires the central density $\rho_0$ and the core radius $r_0$.  These two parameters control the size and shape of the system; for simplicity, we choose the values
\begin{equation}
\begin{array}{lll}
r_0 & = & 500 \, \pc , \\
\rho_0 & = & 30 \, M_\odot \, \pc^{-3} ;\\
\end{array}
\end{equation}
these values are averages of the systems \cite{mar99} used in fitting the adiabatic growth model to a variety of different galaxies.

We carry out the calculation (\ref{eq:time}) numerically, taking the DF dependence on black hole mass to be given by assuming adiabatic growth. This was done for 30 different black hole masses, resulting in an equation of the form 
\begin{equation}
\label{eq:growth}
\frac{\dd M_{bh}}{\dd t} = \kappa M_{bh}^b,
\end{equation}
with $b$ close to 2 for the isothermal system, but rather different in the other cases. Such an equation is readily integrated to give the growth time scale. The mass is not subtracted from the DF itself as both our trials and those of \citet{qui95} suggest that the effect on the results is small, at least for $m_{bh}\le 0.5$.

We first report the calculation for the isothermal sphere (containing an adiabatically grown BH).  Numerical calculation of equation (\ref{eq:time}) gives $b = 2$ and $\kappa = 10^{-18}$ $M_\odot ^{-1}$yrs$^{-1}$; integration of (\ref{eq:growth}) with a seed BH mass of $10^8 M_{\odot}$ then results in a timescale of about nine billion years.  This is close to what is interesting, since this is about the time available since the quasar epoch.  

If we instead use the self-similar DF, modified by the adiabatic growth of a central BH, the timescale becomes absurdly short.  For $\delta = 3/4$, we calculate $b \sim -1/2$ and $\kappa = 10^{42}$ $M_\odot ^{3/2}$yrs$^{-1}$, leading to a very rapid growth (using again a seed mass of $10^8$ $M_\odot$): it would take only $10^{-30}$ years to grow the BH by a factor of ten! The DF for $\delta<1$ increases with increasingly negative energy and the density profile is very steep.  Moreover, the orbits tend to remain rather isotropic as the BH grows, so that essentially all of the mass falls in on a central dynamical timescale.  Setting $\delta > 1$ does slow this growth down substantially; for $\delta = 2$, for example, we get $\kappa = 250$ $M_\odot ^{2/3}$yrs$^{-1}$ with $b = 1/3$, giving a timescale of about 5,000 years.  Increasing $\delta$ increases this time to at most a few hundred thousand years.

The NFW system (modified by the BH growth) is similar in some respects to the self-similar system with $\delta = 2$; both share the same density and radial velocity profiles at small radii.  Indeed, the timescale calculation for the NFW system yields results similar to the $\delta = 2$ case, with the only difference being a smaller value for constant, at $\kappa = 115$ $M_\odot ^{2/3}$yrs$^{-1}$.  This smaller constant gives a timescale of just over 10,000 years for the BH to grow ten times larger than its initial mass of $10^8$ $M_\odot$.  This may be accounted for by the more pronounced anisotropy of the NFW DF at larger radii.

It should be noted that changing the initial constants ($r_0$, $\rho_0$, and the initial BH mass $M_0$) can vary these numbers by a few orders of magnitude in both directions.  From the work of \citet{mar99}, which fits an isothermal sphere to a variety of galaxies, it seems that the core radius can be as small as a few tens of parsecs for ``power-law'' galaxies and as large as a few kiloparsecs for ``core'' galaxies.  Assuming the same scaling laws \citet{mar99} uses, the density $\rho_0$ will depend on the value we take for the core radius.  According to these scaing laws, a radius of $r_0 = 20$ pc, for example, corresponds to a denisty of $\rho_0 = 8360$ $M_\odot$ pc$^{-3}$, while at $r_0 = 2000$ pc the density will be $\rho_0 = 3$ $M_\odot$ pc$^{-3}$.  Using these two sets of values, and assuming that the initial seed BH mass can vary between $10^5$ $M_\odot$ and $10^9$ $M_\odot$, we see that the isothermal sphere can grow by a factor of ten anywhere between a million years and $10^{14}$ years. For the self-similar system with $\delta < 1$, changing the constants has essentially no effect, since the timescale is too short; for $\delta \gg 1$, however, the timescale can approach a few billion years.  The timescale for the NFW system will not go beyond a few hundred million years.

It seems, then, that only the isothermal or Gaussian DF can yield a black hole mass age relation of the type detected by \citet{mer00} by adiabatic growth from a collisionless DF.  The other systems studied here grow more quickly than the isothermal sphere, mainly because of their stronger central density cusp.


\section{THE ADIABATIC $M_{bh} - \sigma$ RELATIONSHIP}

Recent observations \citep{fer00, geb00} have found a strong correlation between the mass of the central BH and the line-of-sight velocity dispersion in the bulge of its host galaxy.  Although various theories have been suggested to explain this relationship (e.g., \citet{hae00}; \citet{ada01}), none have yet to be proven conclusively.  This relationship has been shown to follow
\begin{equation}
\label{eq:mbh-sigma}
M_{bh} \propto \sigma^\alpha,
\end{equation}
where $\alpha$ is somewhere between 3.5 and 5.

The adiabatic growth model has been used by \citet{mar99} to explain various observational properties of black holes, such as the central density cusp and its correlation with the luminosity of the galactic bulge.  His analysis included properties both intrinsic to the adiabatic growth as well as scaling relations based on fundamental-plane-like observations.  We repeat his analysis here to explore the $M_{bh} - \sigma$ relation.

First we calculate any intrinsic relation between the BH mass and the velocity dispersion of the bulge that may arise naturally from the adiabatic growth of the central BH.  Keep in mind, however, that the calculations that follow cannot be compared directly with observations, since they are noise-free and have an infinite resolution.  Regardless, they should give a sense of whether or not the BH growth can give a relation like (\ref{eq:mbh-sigma}).

The line-of-sight velocity dispersion is found by projecting the radial and transverse velocity moments on the plane of the sky.  The velocity moments are calculated from 
\begin{equation}
<V_R^m V_T^n> = \frac{4\pi}{\varrho} \int_{\psi(R)}^{\psi(\infty)} \dd \varepsilon \int_0^{j_{max}} \frac{j \dd j}{r^2 V_R} F(\varepsilon, j) V_R^m V_T^n;
\end{equation}
projecting them on the sky gives
\begin{equation}
<V_P^2>(R_p) = \frac{2}{\Sigma(R_p)} \int_{R_p}^\infty \frac{\dd R \, R \, \rho}{\sqrt{R^2 - R_p^2}} \left[ \left( 1 - \frac{R_p^2}{R^2} \right) <V_R^2> + \frac{R_p^2}{2R^2}<V_T^2> \right],
\end{equation}
where $R_p$ is the projected radius, and $\Sigma$ is the projected density, given by 
\begin{equation}
\Sigma(R_p) = 2 \int_{R_p}^\infty \frac{\dd R \, R \, \rho}{\sqrt{R^2 - R_p^2}}.
\end{equation}

It is a simple matter to predict the dispersion near the centre of the system.  As stated above, the velocity moments simply reflect the Keplerian potential near the BH; thus they take the form
\begin{equation}
<V^2> \propto \frac{m_{bh}}{R},
\end{equation}
and this form is identical regardless of the intial system.  Writing the dispersion as $\sigma = [<V^2>]^{1/2}$, this relation is simply
\begin{equation}
\label{eq:mbh-sigma^2}
m_{bh} = \sigma^2.
\end{equation}

Of course, this relation is applicable only in the innermost regions. Observations are usually done much farther from the centre -- typically near the effective, or half-light, radius of the bulge. This radius corresponds to the core radius $r_0$ for the unperturbed isothermal sphere. To calculate any relationship between the BH mass and the velocity dispersion away from the centre of the system, we must use the DF calculated from the adiabatic growth framework.  

\begin{figure}
\epsscale{0.8}
\plotone{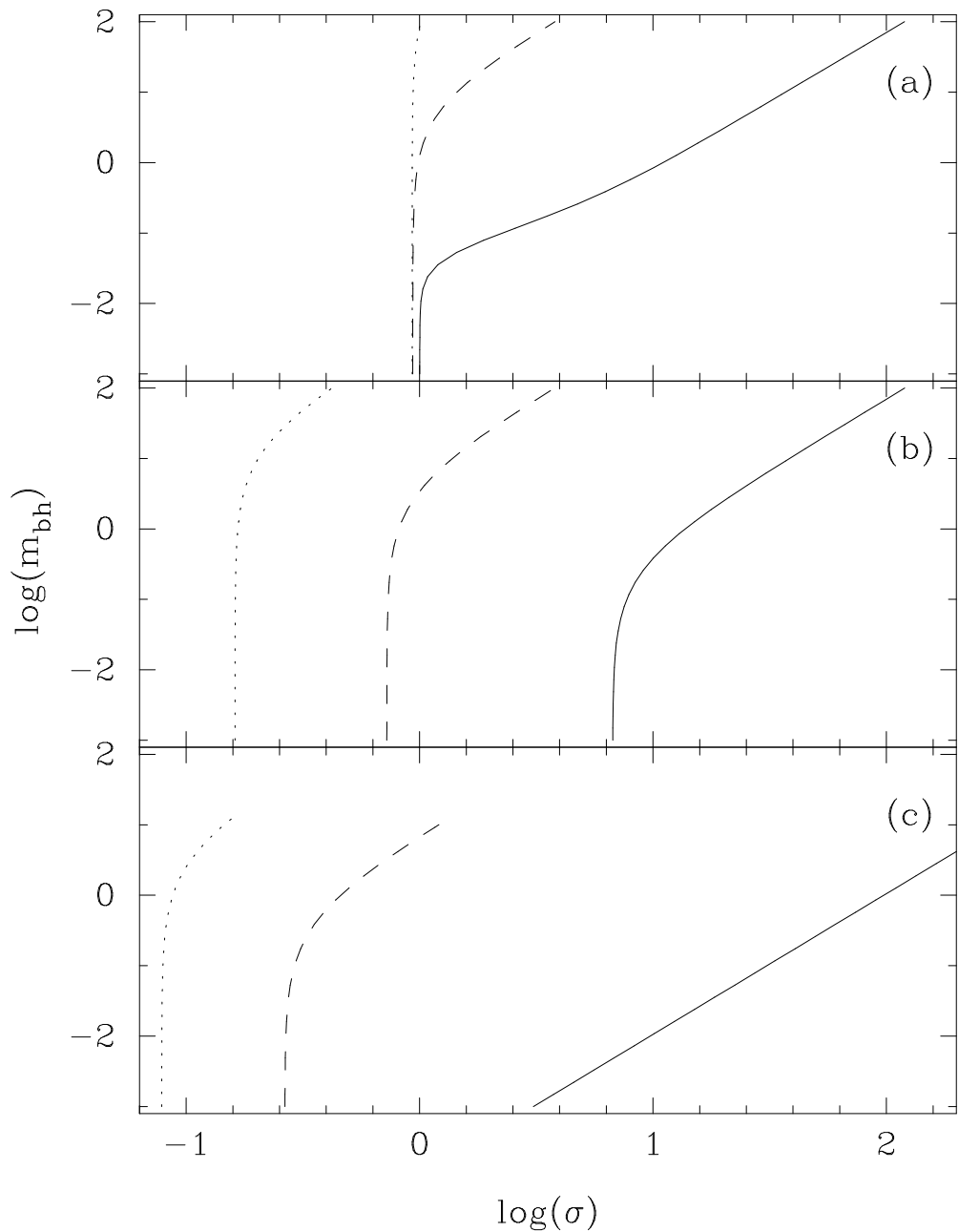}
\caption{$M_{bh}-\sigma$ relationship in the three systems studied.  Panel (a) shows the isothermal sphere, panel (b) is the self-similar system with $\delta = 3/4$, and panel (c) shows the NFW system.  Shown is the relationship at three radii: $R = 10^{-3}$ (solid line), $R = 1$ (dashed line), and $R = 100$ (dotted line).}
\label{fig:mbh-sigma}
\end{figure}

Figure \ref{fig:mbh-sigma} shows the results for the three systems studied, calculated at three different radii:  $R = 10^{-3}$, near the centre; $R = 1$, at the core radius; and $R=100$, well outside of the central region.  The simple predictions made above (\ref{eq:mbh-sigma^2}) are confirmed to exist in this model asymptotically as the mass of the hole becomes large (the solid lines in Figure \ref{fig:mbh-sigma}). As the calculations are done farther out, however, the relation exists only at larger and larger masses (the dashed and dotted lines in Figure \ref{fig:mbh-sigma}). 

Similar calculations for the self-similar and NFW distributions yield almost identical results. The NFW DF is notable for attaining the asymptotic relation at smaller black hole masses than for the isothermal case (the self-similar DF is intermediate), but in all cases no linear relation steeper than that of equation (\ref{eq:mbh-sigma^2}) is found to exist. There is, as can be seen in the figures, a steep shoulder during the approach to the asymptotic limit but this is non-linear and of insignificant extent.

So it seems that the required relation does not arise naturally in the adiabatic growth framework.  This is not suprising, of course; after all, the BH does not create a disturbance much further than the radius $R = m_{bh}$.  Thus only the largest black holes can reach out to the typical radius at which observations are taken, and then they establish a much flatter relation.

In the next section therefore we consider an alternative to adiabatic growth wherein the central black hole and the dark matter halo form together on the dynamical time scale. 

\section{DYNAMICALLY GROWN BLACK HOLES}

The argument in this section is somewhat more speculative than in the preceding sections, and it must ultimately be checked by extensive numerical calculations. However the argument is compelling on dimensional grounds and is consistent with well known solutions and simulations in spherical symmetry \citep{hen99, fil84, ber85}.

We assume that the galaxy forms by the extended collapse of a ``halo'' composed of collisionless matter, and that simultaneously the central black hole is growing proportionally to the halo as matter continues to fall in. We do not assume spherical symmetry. 

The preceding assumption is equivalent to the assumption of multi-dimensional self-similarity as defined in \citep{car91} and in \citet{hen97}. The technique was used in spherical symmetry in \citet{hen95}. The essential idea is that under this assumption one can write the mass inside any surface (the surface in space would be defined by holding $M$ constant at a fixed time) in the halo as 
\begin{equation}
M = \mathcal{M} (\boldsymbol{X}) e^{(3\delta-2\alpha)T}.
\label{eq:mass}
\end{equation}   

Here the vector $\boldsymbol{X}\equiv \boldsymbol{r}e^{-(\delta T)}$ is a scaled position vector and $e^{(\alpha T)}\equiv \alpha t$. The quantities $\delta$ and $\alpha$ are scales in space and time that allow dimensional information to be included in the expressions for the various quantities. Thus the mass scale $\mu$ is determined in terms of $\delta$ and $\alpha$ by the condition that $G$ is a constant of the problem. This yields $\mu=3\delta-2\alpha$ as used above.

The velocity of any particle in the halo may be written consistently as 
\begin{equation}
\boldsymbol{v}=\boldsymbol{Y}e^{(\delta-\alpha)T}.
\label{eq:velocity}
\end{equation}

The quantities $\boldsymbol{X}$ and $\boldsymbol{Y}$ are independent of $T$ during the self-similar collapse and so define a steady-state phase space. Consequently on a fixed spatial surface and averaged over the line of sight rms velocity we can eliminate $T$ between equation (\ref{eq:mass}) and the averaged equation (\ref{eq:velocity}) to obtain 
\begin{equation}
\log{ M}\propto \left(\frac{3\delta/\alpha-2}{\delta/\alpha-1}\right)
\log{\sigma},
\label{eq:Msigma}
\end{equation}
where $\sigma$ is the velocity dispersion along the line of sight. It does not matter which mass surface is chosen in the system if it is truly self-similar of course. Thus the preceding relation applies to the ``bulge'' mass at the bulge scale and to the black hole mass on the black hole scale. However the black hole mass will be simply proportional to the bulge mass during self-similar growth so that \emph{both} masses will obey relation (\ref{eq:Msigma}). This of course also requires that the black hole mass and the bulge mass are also proportional but, given that we are talking about total masses and recalling the vagaries of star formation, this would not necessarily imply a tight bulge luminosity black hole mass correlation.

We may proceed to require that the constant of proportionality in equation (\ref{eq:Msigma}) is equal to the observed \citep{geb00, fer00} constant, say $a$. Then we find that
\begin{equation}
\frac{\delta}{\alpha}=\frac{a-2}{a-3},
\label{eq:DA}
\end{equation}
so that, for example, if $a=4$, then $\delta/\alpha=2$; and if $a=4.5$ then $\delta/\alpha=5/3$. Should $a<4$ (but $>3$ which appears here as a kind of lower permissible limit) say $15/4$, then $\delta/\alpha =7/3$. The principal numerical fact to note is that $\delta/\alpha$ is greater than or less than 2 depending on whether $a$ is less than or greater than 4. 

The reason for the numerical discussion of the preceding paragraph is that the value $\delta/\alpha=2$ is highly significant in spherical models of dark matter halo growth (see e.g. \citet{hen99} for a summary). In these models this ratio is given in terms of the power law index $-\epsilon$ of the initial cosmological density perturbation by 

\begin{equation}
\frac{\delta}{\alpha}=\frac{2}{3}\left(1+\frac{1}{\epsilon}\right).
\label{eq:SSclass}
\end{equation}
Consequently we can infer from equations (\ref{eq:DA},\ref{eq:SSclass}) that the initial cosmological overdensity had the power $-\epsilon$ where 
\begin{equation}
\epsilon=2\left(1-\frac{3}{a}\right).
\label{eq:index}
\end{equation}

If finally we relate $\epsilon$ to the power spectrum index $n$ of the primordial density through the rms profile of such perturbations (other choices are possible, e.g. \citet{hof85}, but similar results are found), then 
\begin{equation}
n=2\epsilon-3.
\end{equation}
Consequently we arrive at a direct link between $a$ and the primordial power spectrum index on the scale of galaxy halos as 
\begin{equation}
n=1-\frac{12}{a}.
\label{eq:Pindex}
\end{equation}

Thus under our interpretation of the black hole mass-velocity-dispersion correlation we are led to conclude that the primordial power spectrum on the scale of galaxies has the index $n$ of equation (\ref{eq:Pindex}). For $a=4$ this yields $n=-2$, while $a=15/4$ and $a=9/2$ yield $n=-11/5$ and $n=-5/3$ respectively. These values for $n$ on the scale of galaxies are all in good agreement with observation, which favours a value near $n=-2$. We conclude that our interpretation of the mass-velocity-dispersion relation as originating in the primordial density profile is consistent with cosmological evidence. 

\section{CONCLUSIONS}

In this paper we have explored the implications of growing a black hole in various dark matter halo distribution functions. Our principal approach was to assume that the black hole grows adiabatically on a time scale long compared with the dynamical time of the halo. The method gives definite predictions for the modified density, radial velocity and anisotropy profiles in the isothermal, self-similar ``power law'' and NFW dark matter distributions. The isothermal calculations reproduced and extended slightly previous work, but the calculations for the self-similar and NFW dark matter halos are new.  Depending on the mass of the black hole the disturbances can be noticeable out to nearly the core radius of the galaxy. Moreover, estimates of black hole growth time scales in the adiabatically modified distribution functions are given. Only in the isothermal DF is there found a reasonable black hole mass-galactic age relation. In no case however does the adiabatic argument give an $M_{bh}-\sigma$ relation that  is close to that observed.

Thus in the concluding section we explored by dimensional argument based on the concept of multidimensional self-similarity \citep{car91,hen97} the possibility that the central black hole grew on a dynamical time scale with the dark matter halo. A simple argument predicts $a=3/(1-\epsilon/2)$ where $a$ is the observed power in the $M_{bh}-\sigma$ relation and $-\epsilon$ is the power in $r$ of the initial cosmolgical density perturbation that produced the galactic halo. Under certain assumptions this power can in turn be related to the power $n$ of the primordial cosmological power spectrum on the scale of galactic halos, so that reasonably, $a=12/(1-n)$. This gives $n=-2$ for $a=4$, which is close to that observed in both cases. We conclude that this suggestion is promising and more work should be done to confirm or infirm the idea that the black hole can grow self-similarly with the dark matter core.   

 
\acknowledgments

This work was supported by the Natural Science and Engineering Research Council of Canada.  We thank the anonymous referee for helpful comments.


\appendix

\section{Algorithm}

This appendix describes the framework that was used to calculate the adiabatic growth models presented above.

We implement this framework using an algorithm similar to Young's (1980), which solves directly for the final state of the system containing the supermassive black hole.  Given a DF that describes the intial system, we
\begin{enumerate}
\item Compute the self-consistent potential $\psi$ and density $\varrho$ for the initial system (eqs. (\ref{eq:density}) and (\ref{eq:potential})).
\item Compute the radial action $i_R(\varepsilon, j)$ (eq. (\ref{eq:radial})) for the intial potential $\psi$.
\item Approximate the final potential $\psi^{*}$ for a given black
  hole mass by
\begin{equation}
\psi^{*} = \psi - \frac{m_{bh}}{R}.
\end{equation}
\item Calculate the radial action $i_R^{*}(\varepsilon, j)$ for this new potential $\psi^{*}$.
\item Equate the intial radial action $i_R(\varepsilon, j)$ and the final action $i_R^{*}(\varepsilon, j)$ to find the energy $\varepsilon$ that has become $\varepsilon^{*}$.  Then the new DF will be 
\begin{equation}
F^{*}(\varepsilon^*, j) = F(\varepsilon (\varepsilon^*,j), j).
\end{equation}
\item Compute a self-consistent density $\varrho^{*}$ and new potential $\psi^{*}$ using $F^{*}$.  Continue back to step 4 and iterate until the density has converged.
\end{enumerate}

The convergence criterion for the algorithm was taken to be when $\varrho^{*}$ changed by less than $10^{-4}$ at all radii.

The computer program developed to implement this algorithm, written in C, uses a radial grid $R_i$, with points spaced logarithmically between an inner radius of $R = 10^{-5}$ and an outer of $R = 10^5$. The various properties of the system are described on this grid; $\varrho_i = \varrho(R_i)$, $\psi_i = \psi(R_i)$, and so on.  The DF is described on a grid of energy and angular momentum points, where the energy points have for convenience the values of the potential on the radial grid, $\varepsilon_i = \psi_i$, and the angular momentum is evaluated as $x = j/j_c$, where $j_c$ is the circular angular momentum and $x$ goes from zero to one.



\newpage
\begin{thebibliography}{}

\bibitem[Adams \emph{et al.}(2001)]{ada01} Adams, F.C., Graff, D.S., and Richstone, D.O.  2001, ApJ 551, L31

\bibitem[Bertschinger(1985)]{ber85} Bertschinger, E.  1985, ApJS, 58, 39

\bibitem[Burkert \& Silk(2001)]{bur01} Burkert, A., \& Silk, J.  2001, ApJ, 554, 151

\bibitem[Carter \& Henriksen(1991)]{car91} Carter, B., \& Henriksen, R.N.  1991, J. Math Phys., 32, 2580

\bibitem[Cipollina \& Bertin(1994)]{cip94} Cipollina, M., and Bertin, G.  1994, A\&A, 288, 43

\bibitem[Ferrarese \& Merritt(2000)]{fer00} Ferrarese, L., \& Merritt, D.  2000, ApJ, 539, L9

\bibitem[Fillmore \& Goldreich(1984)]{fil84} Fillmore, J.A., \& Goldreich, P.  1984, ApJ, 281, 1

\bibitem[Gebhardt \emph{et al.}(2000)]{geb00} Gebhardt, K., \emph{et al.} 2000, ApJ, 539, L13

\bibitem[Gondolo \& Silk(1999)]{gon99} Gondolo, P., and Silk, J.  1999, Phys. Rev. Lett., 83, 1719

\bibitem[Haehnelt \& Kauffmann(2000)]{hae00} Haehnelt, M.G., \& Kauffmann, G.  2000, MNRAS, 318, L35

\bibitem[Henriksen(1997)]{hen97} Henriksen, R.N.  1997, in Scale Invariance and Beyond, Les Houches Workshop, ed. B. Dubrulle, F. Graner, D. Sornette (Berlin: Springer), 63

\bibitem[Henriksen \& Le Delliou(2001)]{hen01} Henriksen, R.N., and Le Delliou, M.  2001, MNRAS, in press

\bibitem[Henriksen \& Widrow(1995)]{hen95} Henriksen, R.N., and Widrow, L.M.  1995, MNRAS, 276, 679

\bibitem[Henriksen \& Widrow(1999)]{hen99} Henriksen, R.N., and Widrow, L.M.  1999, MNRAS, 302, 321

\bibitem[Hoffman \& Shaham(1985)]{hof85}  Hoffman, Y, \& Shaham, J.  1985, ApJ, 297, 16

\bibitem[Lee \& Goodman(1989)]{lee87} Lee, M.H., and Goodman, J.  1989, ApJ, 343, 594

\bibitem[Magorrian \emph{et al.}(1998)]{mag98} Magorrian, J. \emph{et al.}  1998, AJ, 115, 2285

\bibitem[Merrifield \emph{et al.}(2000)]{mer00} Merrifield, M.R., Forbes, D.A., and Terlevich, A.I. 2000, MNRAS, 313, L29

\bibitem[Nakamura(2000)]{nak00} Nakamura, Tadas K.  2000, ApJ, 531, 739

\bibitem[Narayan, Igumenshchev \& Abramowicz(2000)]{nar00} Narayan, R., Igumenshchev, I. V., and Abramowicz, M. A.  2000, ApJ, 539, 798

\bibitem[Navarro, Frenk \& White(1996)]{nav96} Navarro, J.F., Frenk, C.S., and White, S.D.M.  1996, ApJ, 462, 563

\bibitem[Ostriker(2000)]{ost00} Ostriker, J.P.  2000, Phys. Rev. Lett., 84, 5258

\bibitem[Peebles(1972)]{pee72} Peebles, P.J.E.  1972, Gen. Rel. Grav., 3, 61

\bibitem[Quinlan \emph{et al.}(1995)]{qui95} Quinlan, G.D., Hernquist, L., and Sigurdsson, S.  1995, ApJ, 440, 554

\bibitem[Richstone \emph{et al.}(1998)]{ric98} Richstone, D., \emph{et al.} 1998, Nature, 395, A14

\bibitem[van der Marel(1999)]{mar99} van der Marel, R.P.  1999, AJ, 117, 744

\bibitem[Widrow(2000)]{wid00} Widrow, L.M. 2000, ApJS, 131, 29

\bibitem[Young(1980)]{you80} Young, P.  1980, ApJ, 242, 1232

\end{thebibliography}
\end{document}